# Capacity Self-Planning in Small Cell Multi-Tenant 5G Networks

P. Muñoz, O. Sallent, J. Pérez-Romero
Universitat Politècnica de Catalunya (UPC)
[pablo.munoz, sallent, jorperez]@tsc.upc.edu

*Abstract*—Multi-tenancy allows diverse agents sharing the infrastructure in the 5$^{th}$ generation of mobile networks. Such a feature calls for more automated and faster planning procedures in order to adapt the network capacity to the varying traffic demand. To achieve these goals, Small Cells offer network providers more flexible, scalable, and cost-effective solutions compared to macrocell deployments. This paper proposes a novel framework for cell planning in multi-tenant Small Cell networks. In this framework, the tenant's contracted capacity is translated to a set of detailed planning specifications over time and space domains in order to efficiently update the network infrastructure and configuration. Based on this, an algorithm is proposed that considers different actions such as adding/removing channels and adding or relocating small cells. The proposed approach is evaluated considering the deployment of a new tenant, where different sets of planning specifications are tested.

I. INTRODUCTION

The evolution towards the 5$^{th}$ generation (5G) mobile networks requires a revolutionary transformation in order to cope with the increasing traffic demand. One of the main challenges in 5G is the support of miscellaneous services from vertical sectors, which have experienced a large industrial development throughout the last decade [1]. In this context, the multi-tenancy concept allows the sharing of mobile network infrastructure among multiple communication providers, denoted as "tenants", according to specific agreements between the infrastructure provider and each involved tenant.

Sharing the Radio Access Network (RAN) in a flexible way is an important feature for an efficient and cost-effective multi-tenancy implementation. In this respect, the dynamic resource provisioning between tenants has been studied in [2-3], where a central entity is responsible for allocating resources via resource slicing. This kind of solutions are intended for operating in short-term time scales. From a perspective of larger time scales, the multi-tenancy poses unprecedented challenges to the owner of the shared RAN in relation to network planning. Specifically, the management of tenants may involve drastic variations in the network's traffic demand, e.g. due to the aggregation of new tenants. Moreover, the new aggregated traffic is tenant-specific, meaning that some characteristics (e.g. busy hour, type of services, etc.) may substantially differ from tenant to tenant, particularly if these tenants correspond to different vertical sectors. For these reasons, the current framework for network planning needs to be evolved in order to overcome the major challenges in the forthcoming 5G.

A promising solution to cope with the stringent requirements of capacity in 5G is the deployment of Small Cells (SCs), which has been studied in [4-6] with special focus on the cross-tier interference between the macrocell and SCs. An important aspect in SC deployments is that the planning can be made in a more localized way, so that the network is partitioned in small clusters, each of which can be independently planned. The SCs are also more economically attractive and easy-to-deploy nodes than macrocells, so that the SC planning becomes a more flexible process. In addition, network densification is a promising cellular deployment technique to significantly increase the cell-edge user throughput [7-8]. Deploying SCs also has strong implications in the way that the spectrum planning (or channel allocation) is carried out, since the newly deployed SCs will interfere other co-channel SCs. However, the spectrum planning problem can be solved in a more localized fashion than in macrocell deployments, because of the smaller size of SCs and the usage of high carrier frequencies, which facilitate extensive spatial reuse.

Due to the complexity of multi-tenant small cell scenarios, the automation of tasks in network planning is of paramount importance. This automation falls under the so-called *self-planning*, which is defined in [9] as the process of identifying the parameter settings of new network elements, including site locations and hardware configuration. Self-planning was included within the Self-Organizing Network (SON) use cases defined by the Next Generation Mobile Networks (NGMN) alliance [10]. Besides, some commercial planning tools already incorporate certain automated planning capabilities (see e.g. [11]). Furthermore, the consideration of traffic prediction mechanisms enable a more proactive approach, where the need for new SCs deployments and/or reconfigurations can be anticipated in a more automated way.

In view of these challenges, this paper proposes a novel framework for automated planning in multi-tenant networks. The contributions of this paper concentrate on (i) modeling a functional architecture that defines the elements and interactions between the network planning processes and the new multi-tenancy management; (ii) translating the capacity requirements from the Service Level Agreement (SLA) to detailed planning specifications in order to optimize the multi-tenant dimensioning and planning process; (iii) establishing the fundamentals of a new methodology for self-planning in localized areas that exploits the benefits of SCs; and (iv) providing some guidelines for developing a suitable framework

This work has been supported by the EU funded H2020 5G-PPP project SESAME under the grant agreement 671596 and by the Spanish Research Council and FEDER funds under RAMSES grant (ref. TEC2013-41698-R).

for Self-planning in the context of SON.

The remainder of this paper is organized as follows. In Section II the system model is discussed. Section III describes the proposed architecture, identifying functional entities and interfaces. Section IV illustrates the implications of multi-tenancy on network planning through a use case. Finally, the paper summarizes conclusions and identifies future work in Section V.

## II. SYSTEM MODEL

Let us assume a scenario where a certain infrastructure provider owns a RAN comprised of SCs. The SCs are intended to meet the high capacity requirements in localized areas. The provider offers such a RAN to a tenant so that the tenant's customers can get access to the tenant's service. The geographical area of interest is divided into a set $U$ of grid points, called pixels. A subset $U_C \subseteq U$ of these locations are candidate site locations for SCs. Then, let $U_S^{(t)} \subseteq U_C$ be the subset of site locations with deployed SCs at time $t$.

The transmit power and the allocated bandwidth of the $i^{th}$ SC are denoted by $P_i^{(t)}$ and $B_i^{(t)}$, respectively. With respect to the carrier frequency, SCs are assumed to be deployed in higher frequencies than the 1~2 GHz, such as e.g. the 5 GHz considered by the 3GPP as a feasible solution [7]. The frequency band is partitioned into a set $F = \{f_1, ..., f_K\}$ of $K$ orthogonal channels of bandwidth $B$. The subset of channels allocated to SC $i$ at time $t$ is given by $F_i^{(t)} \subseteq F$. Therefore, the total bandwidth allocated to SC $i$ is expressed as $B_i^{(t)} = |F_i^{(t)}| \cdot B$, where $|\cdot|$ denotes cardinality. The capacity of SC $i$ is given by:

$$C_i^{(t)} = B_i^{(t)} \cdot \overline{SE}_i^{(t)}, \qquad (1)$$

where $\overline{SE}_i^{(t)}$ represents the average spectral efficiency achievable at SC $i$. In general, the spectral efficiency depends on the radio access technology and the Signal to Interference plus Noise Ratio (SINR) conditions.

The network infrastructure is shared at time $t$ by a certain number of tenants, denoted by $M^{(t)}$.

## III. FUNCTIONAL ARCHITECTURE

This section focuses on elaborating a reference framework for multi-tenant management from the perspective of network planning. The proposed model is depicted in Fig. 1. The network, represented in the bottom of the figure, is characterized by the network configuration, which is given by $U_S^{(t)}$ and $F_i^{(t)}$. The network can be seen as a source of relevant information for the planning process. In particular, it provides a collection of metrics related to the past and actual traffic demand and also to the quality of the offered services. The information can be given at either the SC-level or pixel-level. In the former case, the metrics are derived from cell counters and they are typically known as Key Performance Indicators (KPIs). In the latter case, the information is derived from cell traces, which contain geo-located measurements from users.

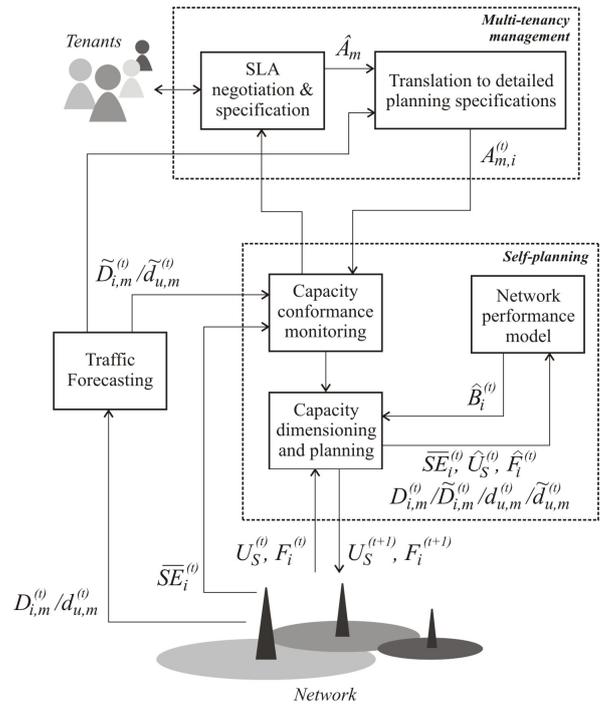

Fig. 1. Functional architecture.

Let denote as $D_{i,m}^{(t)}$ the traffic demand (in Mbps) of tenant $m$ in SC $i$ at time $t$, calculated as the sum of the traffic from all the users attached to SC $i$. The tenants' traffic $D_{i,m}^{(t)}$ is aggregated into a metric, $D_i^{(t)}$, which provides the total traffic demand in SC $i$, calculated as:

$$D_i^{(t)} = \sum_{m=1}^{M^{(t)}} D_{i,m}^{(t)}. \qquad (2)$$

Let also $d_{u,m}^{(t)}$ be the traffic demand (in Mbps) of tenant $m$ in the $u^{th}$ pixel of the scenario ($u \in U$). The metric $d_u^{(t)}$ is computed in a similar way to (2) in order to determine the total traffic at the pixel-level.

The functional architecture of Fig. 1 includes two main entities described in the following. These entities can be part of the management systems such as the Element Manager (EM) or the Network Manager (NM) [12].

### A. Multi-tenancy management entity

The multi-tenancy management entity acts as an interface between the tenants and the network planning activities of the network provider. From the perspective of planning, the SLA defines the contracted capacity $\hat{A}_m$ (in Mbps) that tenant $m$ demands to the network provider. Normally, it is expressed in terms of aggregate (or average) values over relatively coarse time and space scales. The SLA may also include some other guarantees, e.g. related to QoS metrics.

For network planning purposes, the SLA has to be expressed in smaller time and space scales that can be more easily used when taking planning decisions. In particular, the contracted capacity $\hat{A}_m$ is translated into a set of detailed planning specifications $A_{m,i}^{(t)}$ that depend on SC $i$ and time $t$. To do this, the current or predicted traffic demand in the network can be employed. This process, which ensures that the contracted capacity is provided, depends exclusively on the network provider's side. As a consequence, the SLA is simplified and tenants are excluded from gaining a detailed picture of the network infrastructure.

*B. Self-planning entity*

According to Fig. 1, the detailed planning specifications, $A_{m,i}^{(t)}$, are used by the self-planning entity, whose aim is twofold. The first objective is to check whether or not the deployed network capacity fits the tenants' demand. The second is to provide the required changes in the network layout and channel allocation, given by $U_S^{(t+1)}$ and $F_i^{(t+1)}$ respectively, in case there is a lack of capacity. The self-planning entity follows an automated approach characterized by running an iterative process that is executed during the network operation assuming that a set of SCs have already been deployed. Thus, the currently deployed infrastructure is incrementally adapted to the evolving tenants' requirements to make capacity expansion smoother, less costly and faster. The proposed approach is applied to smaller regions that are covered by a subset (or cluster) of SCs, so that dimensioning and planning tasks are accelerated and simplified. As a result, dimensioning and planning can be regarded as an automated function that can be easily integrated into the SON framework. Both traditional and SON-based planning are complementary approaches to drive network expansion at different time scales.

The *capacity conformance monitoring* module watches over the network to determine when the network infrastructure has to be reconfigured in order to meet the tenant's traffic demand while minimizing over-provisioning. The required bandwidth in the SCs can vary due to high tenant's actual traffic demand $D_{i,m}^{(t)}$, the addition or removal of new tenants (i.e. variations in $M^{(t)}$) or changes in the contracted capacity $A_{m,i}^{(t)}$. Also, if the process is executed proactively, the predicted traffic demand can be considered. To this end, the *traffic forecasting* entity provides the predicted traffic demand in time $t$ at the SC-level (which can be computed from historical data using statistical models) as input in the self-planning entity. The proactive response is key as long as the deployment of new infrastructure may require substantial time compared to the evolution of the traffic demand.

The capacity conformance is conducted in terms of the required bandwidth $\tilde{B}_i^{(t)}$ by SC $i$, which can be estimated as:

$$\tilde{B}_i^{(t)} = \frac{1}{SE_i^{(t)}} \sum_{m=1}^{M^{(t)}} \min\left(D_{i,m}^{(t)}, A_{m,i}^{(t)}\right). \quad (3)$$

If the traffic demand of tenant $m$ at SC $i$ is below the SLA's planning specification, $D_{i,m}^{(t)}$ is used to provide cost-effective dimensioning, since the SC's bandwidth would fit the actual required bandwidth. If, on the contrary, the traffic demand exceeds the SLA's planning specification, the required bandwidth is then limited by $A_{m,i}^{(t)}$. The bandwidth of SC $i$, $B_i^{(t)}$, is dimensioned so that the required bandwidth $\tilde{B}_i^{(t)}$ is satisfied at the busy hour $t_B$, which is calculated as:

$$t_B = \arg\max_\tau \left(\tilde{B}_i^{(\tau)}\right), \quad \tau = t - T + 1, ..., t. \quad (4)$$

According to this, the *capacity conformance monitoring* module triggers the *capacity dimensioning and planning* module if the following condition is fulfilled for any of the deployed SCs in $L$ consecutive time steps:

$$\tilde{B}_i^{(t_B)} > \alpha \cdot \left|F_i^{(t_B)}\right| \cdot B \quad (5)$$

where $\alpha \in [0,1]$ is an adjustable parameter.

Lastly, note that, if the total traffic demand of any tenant exceeds the contracted capacity, the *capacity conformance monitoring* module should communicate the multi-tenancy management entity the need of reviewing (or negotiating) the SLA in order to meet the traffic demand.

The *capacity dimensioning and planning* module aims to determine the optimal solution (i.e. an updated RAN) to cope with the varying traffic demand. A candidate solution is represented by $\hat{U}_S^{(t)}$ and $\hat{F}_i^{(t)}$, which represent a modified version of the actual network deployment and spectrum allocation, respectively. The required bandwidth of the candidate solution, $\hat{B}_i^{(t)}$, is obtained from the *network performance model*, which emulates the behavior of the network with a certain layout and configuration. Unlike the SC bandwidth (measured in steps of $B$ MHz), $\hat{B}_i^{(t)}$ is a continuous variable that depends on the traffic demand and the spectral efficiency.

The dimensioning and planning is modeled as an iterative process, initiated after satisfying (5), where a set of conditions are sequentially checked at each time step in order to trigger specific planning actions (i.e. adding/removing a channel and deploying/relocating a SC). Such actions are accumulated during the planning process and, after that, the infrastructure provider is notified about the changes in the network to be implemented. The planning process is summarized in Algorithm 1, where $N_{max}^{SC}$ is the maximum number of SCs that can be deployed in the area of interest, $K_{max}$ is the maximum number of channels that can be allocated in a SC and $\beta, \gamma \in [0,1]$ are adjustable parameters. In detail, in steps 1-6, the planning process focuses on extending the capacity in areas with a lack of capacity, while in steps 7 to 12, this process aims at minimizing the capacity overprovisioning. During the execution of this process, a certain planning action (e.g. adding a channel) can be canceled due to the execution of the opposite action (e.g. removing a channel) depending on the actions carried out between the two (e.g. a channel added in step 2 may no longer be needed if a SC is later on added in step 5). In step 2, when a new channel has to be added to SC $j$ the channel is selected so that the SC-to-SC distance between SC $j$ and the closest neighboring SC using the same channel is the maximum possible. When a planning action is selected in either step 2, 5, 8 or 11, the network performance model is launched to obtain the value of $\hat{B}_i^{(t)}$ corresponding to the new network configuration. In case a new SC has to be deployed, $\hat{B}_i^{(t)}$ is calculated for each candidate site in the area of interest. Then, the site with the lowest required bandwidth is selected.

In order to select the optimal location of a SC, Algorithm 1 is based on exhaustive search. Since the search space is assumed to be a relatively small geographical area, the computational complexity of this method does not require sophisticated combinatory optimization.

| | **Algorithm 1** Capacity dimensioning and planning |
|---|---|
| 1: | **While** $\exists j \mid \hat{B}_j^{(t_B)} > \alpha \cdot \left| \hat{F}_j^{(t)} \right| \cdot B$ **and** $\left| \hat{F}_j^{(t)} \right| < K_{max}$ |
| 2: | Add a new channel to SC $j$ |
| 3: | **End** |
| 4: | **While** $\exists j \mid \hat{B}_j^{(t_B)} > B \cdot \dfrac{\left| U_S^{(t)} \right|}{N_{max}^{SC} / K_{max}}$ **and** $\left| U_S^{(t)} \right| < N_{max}^{SC}$ |
| 5: | Add a new SC with objective: $\min \sum_{i \in \hat{U}_S^{(t)}} \hat{B}_i^{(t_B)}$ |
| 6: | **End** |
| 7: | **While** $\exists j \mid \hat{B}_j^{(t_B)} < \beta \cdot \left( \left| F_j^{(t)} \right| - 1 \right) \cdot B$ **and** $\left| \hat{F}_j^{(t)} \right| > 1$ |
| 8: | Remove a channel from SC $j$ |
| 9: | **End** |
| 10: | **While** $\exists j \mid \hat{B}_j^{(t_B)} < \gamma \cdot B$ |
| 11: | Remove SC $j$ |
| 12: | **End** |

## IV. USE CASE: ADDING A NEW TENANT

Let us suppose that the network is being shared among several tenants and that, at a certain time $t$, the infrastructure provider adds a new tenant $m$, after negotiating a certain contracted capacity $\hat{A}_m$. Based on this, the detailed planning specifications $A_{m,i}^{(t)}$ must be generated so that the capacity conformance monitoring module can determine whether any change/update of the network configuration is required or not. Specifically, $A_{m,i}^{(t)}$ is calculated based on the temporal and spatial variations of the traffic demand. The temporal variation of the traffic demand is mainly given by the traffic fluctuations that take place over one day's time. Such a temporal pattern is typically repeated over different days. Based on this, let $A_m^{(t_B)}$ be the detailed planning specification at the busy hour, which can be estimated from $\hat{A}_m$ and from the time variations of the other tenants' traffic demand.

Regarding the spatial variations of the traffic demand, the contracted capacity at the busy hour $A_m^{(t_B)}$ is distributed among the number $\left| U_S^{(t)} \right|$ of deployed SCs taking into account the following condition:

$$A_m^{(t_B)} = \sum_{i \in U_S^{(t)}} A_{m,i}^{(t_B)},  \quad (6)$$

where $A_{m,i}^{(t_B)}$ is the contribution of the contracted capacity in SC $i$. Depending on the spatial correlation that can be expected between the tenant's traffic demand and the actual network's traffic demand, the detailed planning specifications per cell for tenant $m$ can be formulated in different ways:

- *Uniform distribution*. In case that the spatial traffic demand of the new tenant is unknown, an even distribution among the SCs is assumed. Estimation can be conducted at the SC-level (7) or pixel-level (8):

$$A_{m,i}^{(t_B)} = \dfrac{A_m^{(t_B)}}{\left| U_S^{(t)} \right|}, \quad (7)$$

$$A_{m,i,u}^{(t_B)} = \dfrac{A_m^{(t_B)}}{\left| U^{(t)} \right|}, \quad (8)$$

where $A_{m,i,u}^{(t_B)}$ stands for the contracted capacity at the $u^{th}$ pixel that is served by SC $i$ and $\left| U^{(t)} \right|$ is the total number of pixels in the area.

- *Correlated distribution*. In case that correlation between the traffic demand for the new tenant and the already existing tenants is expected, areas with higher traffic demand of other tenants will receive a greater contribution of $A_m^{(t_B)}$. Such an estimation can also be conducted at either SC- or pixel-level. In the former case, using the information on KPIs that measure $D_i^{(t_B)}$ in SC $i$ as an estimation of the spatial traffic demand of tenant $m$, the detailed planning specification is given by:

$$A_{m,i}^{(t_B)} = A_m^{(t_B)} \cdot \dfrac{D_i^{(t_B)}}{\sum_{p \in U_S^{(t)}} D_p^{(t_B)}}. \quad (9)$$

In the latter case, the traffic measurements at the pixel-level are taken from cell traces that provide geo-located information for each user in an automatic way. Thus, the specification is calculated as:

$$A_{m,i,u}^{(t_B)} = A_m^{(t_B)} \cdot \dfrac{d_u^{(t_B)}}{\sum_{v \in U^{(t)}} d_v^{(t_B)}}, \quad (10)$$

where $d_u^{(t_B)}$ is the total traffic demand in the $u^{th}$ pixel of the scenario.

### A. Simulation scenario

An urban SC scenario with dimensions 0.2 km × 0.2 km and a grid resolution of 3 m has been considered. To represent the areas where deploying SCs is not possible, e.g. because of backhaul and site acquisition constraints, 2% of the points (or pixels) in the scenario have been randomly selected as candidate site locations. The network layout and the traffic demand at the busy hour in the situation before the consideration of the new tenant are represented in Fig. 2(a), where the triangles represent the location of the deployed SCs and the values in brackets are the number of allocated channels. It is observed that 4 SCs have been deployed to meet a total traffic demand of 86.5 Mbps, which is non-uniformly distributed over the considered area. Specifically, the traffic demands supported by SCs 1-4 before the new tenant's arrival are 22.5, 27.9, 19.3 and 16.6 Mbps, respectively.

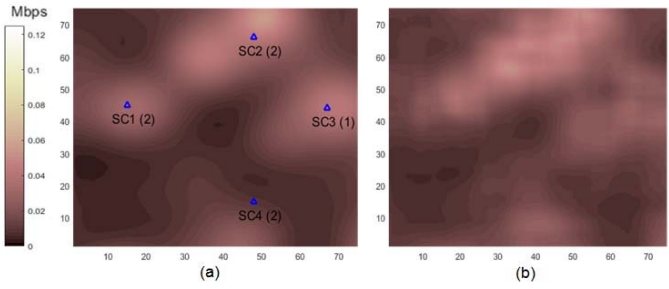

Fig. 2. (a) Traffic demand and network deployment in the initial situation (before new tenant's arrival); (b) Traffic demand of the new tenant.

Regarding the network performance model, each pixel is served by the SC that provides the highest received power. The path loss is computed using the ITU InH model in [13]. The carrier frequency is 5 GHz. The frequency band is composed of 4 channels, with channel bandwidth $B = 20$ MHz. The SCs are assumed to be omnidirectional with an antenna gain of 2 dB. The transmit power $P_i^{(t)}$, which can vary between 24 and 10 dBm, is configured for each SC to have a SINR of 9 dB at $\frac{\sqrt{3}}{2}$ of the inter-site distance [7]. The terminal noise figure is 9 dB. The spectral efficiency function $SE(SINR)$ used to compute the average spectral efficiency $\overline{SE}_i^{(t)}$ at SC $i$ depending on the SINR at each pixel is obtained from Section A.1 in [14] with $SE_{max} = 4.4$ b/s/Hz.

From a perspective of planning, the parameters used in the capacity conformance monitoring module to trigger the planning actions are configured as follows: $\alpha = 0.9$, $\beta = 0.7$ and $\gamma = 0.05$. In addition, the maximum number of allocated channels per SC is set to $K_{max} = 2$, while the maximum number of SCs that can be deployed in the considered area is set to $N_{max}^{SC} = 10$.

Let assume that the SLA of the new tenant is translated to a specification at the busy hour of $A_m^{(t_B)} = 100$ Mbps and that the traffic demand at the busy hour is spatially distributed as shown in Fig. 2(b). It can be observed that the new tenant's spatial traffic demand exhibits quite high correlation with already existing tenants.

After generating the detailed planning specifications, it is observed that in the capacity conformance monitoring module condition (5) is satisfied for the four deployed SCs, so that the capacity dimensioning and planning module is launched.

### B. Analysis of the network planning solutions

Fig. 3(a-e) show the results of the planning process for different sets of detailed planning specifications. In the approaches shown in Fig. 3(a-d), the new tenant's spatial traffic demand distribution is assumed to be unknown, and the planning is then carried out using the detailed planning specifications from the methods explained in Section IV. In particular, the total traffic demand in the figures is calculated as the actual traffic demand from existing tenants plus the estimated new tenant's demand.

For the uniform distribution at the SC-level method, Fig. 3(a) shows that, given that SC 2 and 3 have the smallest service areas, the traffic demand per pixel in these SCs is higher than in SC 1 and 4. As a result, the planning strategy has added 4 new SCs, and three of them (SC6, SC7 and SC8) are located close to the existing SC2 and SC3 so that they can offload traffic from these cells. For the correlated distribution at the SC-level method, represented in Fig. 3(b), it is observed that SCs 1 and 2, which initially carried more traffic (22.5 and 27.9 Mbps respectively), receive proportionally more traffic from the new tenant than other SCs. Compared to the previous method, it is observed that an additional cell (SC 9) is located closer to SC 3, which satisfied the condition to deploy a new SC in step 4 of Algorithm 1. The method based on uniform distribution at the pixel-level, illustrated in Fig. 3(c), reveals that a greater number of SCs is needed to meet the expected traffic demand. The last method [see Fig. 3(d)], based on correlated distribution at the pixel-level, produces the largest variations in the traffic demand per pixel. In this case, the planning strategy has added 4 new SCs, placing one of them (SC8) in an area with high traffic density. Moreover, this method provides a lower number of allocated channels than the method based on uniform distribution at SC level of Fig. 3(a), although they have the same number of deployed SCs.

As a reference for comparison, Fig. 3(e) shows the network planning considering that the actual spatial traffic distribution of the new tenant is known (i.e., the planning is conducted considering the actual traffic of the existing tenants plus the real traffic that the new tenant will offer). In line with the methods based on correlated distribution, a SC (in this case, SC8) has been placed in one of the areas with high traffic density. This differs from the methods based on uniform distribution, which are not able to correctly place SC7 in such an area of high traffic density [see Fig. 3(a) and (b)]. In addition, the network layout in the reference case of Fig. 3(e) is very similar to that based on correlated distribution at the pixel-level of Fig. 3(d), which differs only in the location of SC 8.

### C. Analysis of the network operation with the new tenant

This section evaluates the solutions of the planning algorithm when the new tenant's service is operative and the actual traffic demand of the new tenant is as considered in Fig. 2(b). For each method, it is assumed that the network has been deployed as dictated by the planning (i.e. the real network layouts are as illustrated in Fig. 3(a-e)). Table I shows the required bandwidth $\tilde{B}_i^{(t)}$ in each SC considering the actual traffic demand. The last row in the table shows values aggregated over all the SCs. The reference scheme requires the lowest number of SCs (i.e. it minimizes deployment costs) and bandwidth, which is expected because it considers the real traffic distribution. The methods based on correlated distribution at the pixel-level and on uniform distribution at SC-level also achieve a deployment with only 8 SCs, while the former requires a lower value of required bandwidth because its network layout fits better the traffic demand (i.e. the solution is more spectrally efficient). In this respect, it can be observed that the required bandwidth with this method is very close to that required by the reference scheme. This reflects the superior performance of the method based on correlated distribution at the pixel-level in the considered scenario.

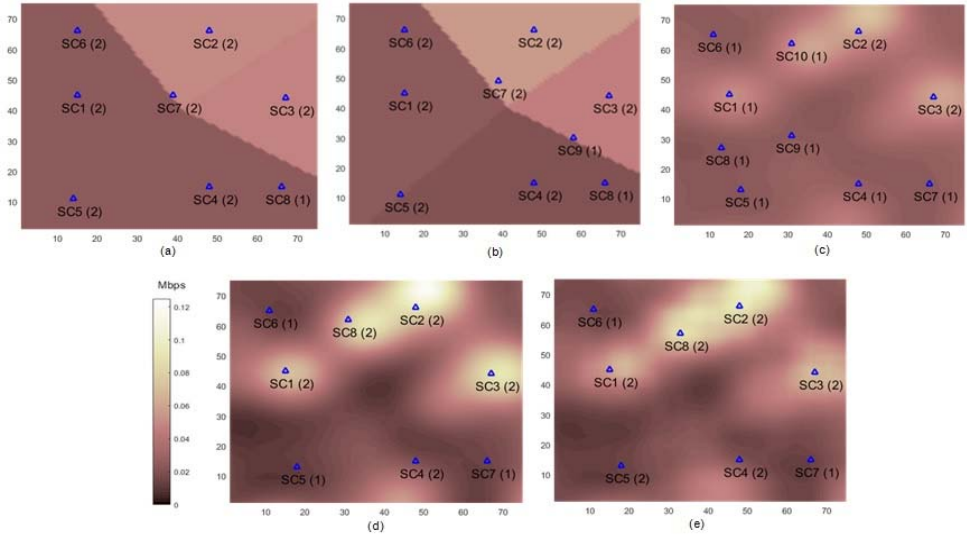

Fig. 3. Network deployment and estimated traffic demand using the detailed planning specifications: (a) Based on uniform distribution at the SC-level, (b) Based on correlated distribution at the SC-level, (c) Based on uniform distribution at the pixel-level, and (d) Based on correlated distribution at the pixel-level; (e) Network deployment with real traffic demand.

## V. CONCLUSION

This paper has focused on the cell planning problem for multi-tenant networks and SC deployments. The problem has been modeled as an iterative process, following a SON approach, where a condition to detect capacity issues is periodically checked in order to automatically trigger specific planning actions, such as adding/removing a channel or deploying/relocating a SC.

The approach has been illustrated through an example use case where a new tenant is added to the network. In particular, a set of detailed planning specifications in the time and space domains has been generated using the actual traffic demand and network layout information. Results show that capacity overprovisioning can be minimized by reducing uncertainties about the spatial and temporal correlations between the new tenant's traffic and the actual traffic in the network. In this way, the most adequate planning specifications can be selected based on this correlation in order to provide efficient planning.

As future work, it is planned to further analyze the planning methodology with focus on the planning actions and their triggering conditions. In particular, the adjustable parameters in these conditions can be tuned to provide a more reactive or proactive behavior in the cell planning.

TABLE I. REQUIRED BANDWIDTH [MHZ] FOR EACH SC

| SC | REFERENCE (ACTUAL TRAFFIC KNOWN) | UNIFORM SC-LEVEL | CORR. SC-LEVEL | UNIFORM PX-LEVEL | CORR. PX-LEVEL |
|---|---|---|---|---|---|
| 1 | 20.9 | 17.7 | 16.7 | 9.3 | 20.3 |
| 2 | 20.7 | 29.9 | 27.7 | 21.7 | 21.3 |
| 3 | 23.3 | 21.6 | 14.7 | 19.2 | 23.4 |
| 4 | 13.7 | 16.3 | 16.8 | 14.6 | 14.6 |
| 5 | 12.5 | 11.6 | 11.9 | 10.5 | 19.7 |
| 6 | 5.5 | 12.2 | 10.3 | 4.2 | 4.1 |
| 7 | 10.5 | 20.3 | 18.3 | 10.3 | 10.4 |
| 8 | 21.5 | 10.7 | 5.9 | 2.5 | 20.0 |
| 9 | - | - | - | 10.1 | 10.3 |
| 10 | - | - | - | - | 13.7 |
| tot | 128 | 140 | 132 | 116 | 133 |